Suprathermal Ion Energy spectra and Anisotropies near the Heliospheric Current Sheet crossing observed by the Parker Solar Probe during Encounter 7

M. I. Desai[1,2], D. G. Mitchell[3], D. J. McComas[4], J. F. Drake[5], T. Phan[6], J. R. Szalay[4], E. C. Roelof[3], J. Giacalone[7], M. E. Hill[3], E. R. Christian[8], N. A. Schwadron[9], R. L. McNutt Jr.[3], M. E. Wiedenbeck[10], C. Joyce[4], C. M. S. Cohen[10], A. J. Davis[10], S. M. Krimigis[3], R. A. Leske[10], W. H. Matthaeus[11], O. Malandraki[12], R. A. Mewaldt[10], A. Labrador[10], E. C. Stone[10], S. D. Bale[6], J. Verniero[8], A. Rahmati[6], P. Whittlesey[6], R. Livi[6], D. Larson[6], M. Pulupa[6], R. J. MacDowall[8], J. T. Niehof[9], J. C. Kasper[13], and T. S. Horbury[14]

[1]Southwest Research Institute, 6220 Culebra Road, San Antonio, TX 78238, USA

[2]Department of Physics and Astronomy, University of Texas at San Antonio, San Antonio, TX 78249, USA

[3]Johns Hopkins University/Applied Physics Laboratory, Laurel, MD 20723, USA

[4]Department of Astrophysical Sciences, Princeton University, NJ 08544, USA

[5]Department of Physics, Institute for Physical Science and Technology, University of Maryland, College Park, MD 20742, USA

[6]Physics Department and Space Sciences Laboratory, University of California, Berkeley, CA

[7]The University of Arizona, Lunar and Planetary Laboratory, Tucson, AZ 85721, USA

[8]NASA Goddard Space Flight Center, Greenbelt, MD 20771, USA

[9]University of New Hampshire, 8 College Road, Durham NH 03824, USA

[10]California Institute of Technology, Pasadena, CA 91125, USA

[11]University of Delaware, Newark, DE 19716, USA

[12]Institute for Astronomy, Astrophysics, Space Applications & Remote Sensing (IAASARS), National Observatory of Athens, Greece

[13]University of Michigan, Ann Arbor, MI 48109, USA

[14]Faculty of Natural Sciences, Department of Physics, Imperial College, London, UK

## ABSTRACT

We present observations of $\gtrsim$10-100 keV nucleon$^{-1}$ suprathermal (ST) H, He, O, and Fe ions associated with crossings of the heliospheric current sheet (HCS) at radial distances <0.1 au from the Sun. Our key findings are: 1) very few heavy ions are detected during the 1$^{st}$ full crossing, the heavy ion intensities are reduced during the 2$^{nd}$ partial crossing and peak just after the 2$^{nd}$ crossing; 2) ion arrival times exhibit no velocity dispersion; 3) He pitch-angle distributions track the magnetic field polarity reversal and show up to ~10:1 anti-sunward, field-aligned flows and beams closer to the HCS that become nearly isotropic further from the HCS; 4) the He spectrum steepens either side of the HCS and the He, O, and Fe spectra exhibit power-laws of the form ~E$^{-4-6}$; and 5) maximum energies $E_X$ increase with the ion's charge-to-mass (Q/M) ratio as $E_X/E_H \propto (Q_X/M_X)^{\propto}$ where $\propto$~0.65—0.76, assuming that the average Q-states are similar to those measured in gradual and impulsive solar energetic particle events at 1 au. The absence of velocity dispersion in combination with strong field-aligned anisotropies closer to the HCS appears to rule out solar flares and near-sun coronal mass ejection-driven shocks. These new observations present challenges not only for mechanisms that employ direct parallel electric fields and organize maximum energies according to E/Q, but also for local diffusive and magnetic reconnection-driven acceleration models. Re-evaluation of our current understanding of the production and transport of energetic ions is necessary to understand this near-solar, current-sheet-associated population of ST ions.

## 1. INTRODUCTION

Observations over the last two decades have provided compelling evidence that suprathermal (ST) ions in the energy range between ~1.5-2 times the bulk solar wind (SW) energy up to ~100 keV nucleon$^{-1}$ are an important constituent of the source population that gets accelerated to higher energies via a wide variety of heliospheric phenomena (e.g., see the review by Desai & Giacalone 2016). Examples include coronal mass ejection- or CME-driven large gradual solar energetic particle events (LSEPs; Desai et al. 2006), interplanetary (IP) shock-associated energetic storm particle events (ESPs; Desai et al. 2001, 2003), corotating interaction region–associated particle events (CIRs; Mason et al. 2012a; Allen et al. 2019), and the Earth's bow shock (e.g., Dwyer et al., 1997; Starkey et al., 2021). With contributions from multiple sources such as the heated SW, impulsive and gradual SEP events, IP shock-associated ESPs, as well as interstellar and inner source pick-up ions, the pool of suprathermal material inside 1 au is highly dynamic and varies in time and location (e.g., Mason et al. 2006; Dayeh et al. 2017). Thus, research devoted to understanding the origin and acceleration of ST ions has gained significant importance as it constitutes a key to unlocking details of the physics of particle acceleration throughout the heliosphere and beyond (e.g., Caprioli et al. 2018).

On 2021 January 17, when Parker Solar Probe (PSP) was close to perihelion at a radial distance of <0.1 au at ~20 R$_s$ during orbit 7 (E07), the Integrated Science Investigation of the Sun-Energetic Particle-Lo instrument (IS ☉ IS/EPI-Lo; McComas et al. 2016) observed significant increases in the intensities of <100 keV nucleon$^{-1}$ ST protons and heavy ions, such as He, O, and Fe both before and after two distinct crossings of the heliospheric current sheet (HCS) as identified by the magnetic field and SW plasma instruments. In this paper, we present a detailed analysis of the properties of ST ions associated with this HCS crossing, particularly focusing on He time-intensity profiles, ion arrival times and velocity dispersion, time evolution of He pitch-angle distributions (PADs) and anisotropies, and ion energy spectra and maximum observed energies. We discuss implications of these results in terms of potential sources, expectations of remote vs. local acceleration mechanisms, and existing propagation models of ST ions in the inner heliosphere.

## 2. INSTRUMENTATION & DATA ANALYSIS

Since its launch on 2018 August 12, NASA's Parker Solar Probe (PSP, Fox et al., 2016) has completed its first nine orbits around our Sun and provided unprecedented observations of our inner heliosphere. PSP will use three more Venus flybys to reach within ~0.04 au, i.e., ~9 solar radii ($R_s$) of the Sun's surface. PSP carries four instrument suites, namely, 1) the Solar Wind Electrons, Alphas, and Protons Investigation (SWEAP; Kasper et al. 2016); 2) the Electromagnetic Fields Investigation (FIELDS; Bale et al. 2016); 3) IS☉IS (McComas et al. 2016); and 4) the Wide field Imager for Solar Probe (WISPR; Vourlidas et al. 2016). Together these suites provide comprehensive measurements of the in-situ SW, the electromagnetic fields, and energetic particles, as well as contextual white light images of the solar corona and the inner heliosphere.

The IS☉IS suite comprises two instruments – the low-energy Energetic Particle Instrument – EPI-Lo and the high-energy Energetic Particle Instrument – EPI-Hi (McComas et al. 2016) to provide detailed measurements of the energy spectra, arrival directions, and composition of H–Fe ions from ~0.02 up to 200 MeV nucleon$^{-1}$ as well as of the energy spectra and arrival directions of 0.025 to 6 MeV electrons (e.g., McComas et al., 2018). We also use supporting data from the FIELDS and SWEAP instrument suites.

In this study, our primary data is taken from the IS☉IS/EPI-Lo instrument (McComas et al. 2016; Hill et al. 2017). EPI-Lo measures the energy spectra, composition, and arrival directions of ions above ~0.02 MeV nucleon$^{-1}$ that enter one of its 80 apertures. As the ions pass through the start foils located at the entrance of each aperture, they generate secondary electrons which provide the start time-of-flight (ToF) signal. The ions then strike a stop foil where they generate secondary electrons for the stop ToF signal. Above specific energies for each species, a solid-sate detector (SSD) directly behind the stop foil also measures the residual kinetic energy (E) of the ion. The coincidence between start, stop, and energy signals provide triple coincidence measurements, and the measured ToF and E are used to identify the species. All valid triple coincidence ion events are binned according to the identified species in pre-defined incident energy ranges by onboard lookup tables based on pre-flight calibration data (see McComas et al. 2016) and on the instrument response during the 2020 November CME-associated SEP/ESP event when count rates for all species were sufficiently high to populate the corresponding tracks in the ToF vs. E matrix (see Hill et al. 2021, in preparation). In this study we use the binned "ion composition" ToF vs. E and the raw triple coincidence H, He, O, and Fe data directly from the ToF vs. E matrices for characterizing the properties of ST ions associated with the HCS crossing.

## 3. OBSERVATIONS

### 3.1 Overview and Time-Histories

Figure 1 provides an overview of the IS⊙IS, SWEAP, and FIELDS observations during the 18-hour interval starting from 0600 UT on 2021 January 17 and ending at 0000 UT on 2021 January 18. During this period, IS⊙IS observes a population of dispersionless suprathermal ions (i.e., simultaneous intensity increases at all measured energies) that is likely associated with crossing of the HCS at ~1340 UT on 2021 January 17. The figure shows the (a) ~15-90 keV nucleon$^{-1}$ H, He, O, and Fe omni-directional intensities, (b) ~30-85 keV nucleon$^{-1}$ He omni-directional intensities, (c) ~30-90 keV nucleon$^{-1}$ He spectral slope, (d) 1/ion speed spectrograms constructed using the ToF vs. E counts for all species, (e) ~35-70 keV nucleon$^{-1}$ He PADs, (f) ~253-314 eV electron PADs – the field-aligned electron strahl – measured by SWEAP, (g) bulk SW speed, and the interplanetary magnetic field or IMF magnitude (h) and radial component (i). Blue vertical lines identify the following five time intervals, as identified from the FIELDS and SWEAP data: (1) prior to the 1$^{st}$ crossing, (2) during the 1$^{st}$ full crossing, (3) in between the two crossings, (4) during the 2$^{nd}$ partial crossing, and (5) after the 2$^{nd}$ crossing. These different regions can also be seen from the magnetic field magnitude and the radial component in Figures 1h & 1i.

Figures 1a & 1b show that IS☉IS/EPI-Lo observed significant increases in <100 keV/n H, He, O, and Fe intensities both before and after the two HCS crossings. Very few heavy ions are detected during the 1$^{st}$ full crossing while the intensities are reduced during the 2$^{nd}$ partial crossing. The He intensities peak just after the magnetic field polarity reversal associated with the 2$^{nd}$ partial crossing, as seen in Figure 1i. Figure 1c shows that the He spectral slope varies between ~4-8 before the 1$^{st}$ full crossing, and remains relatively constant at ~6 after the 2$^{nd}$ HCS crossing (i.e., during the peak intensity enhancement seen in Figures 1a & 1b). Figures 1e and 1f show that the ST He PADs and the electron strahl track the magnetic field polarity reversal associated with the HCS, and show an anti-sunward anisotropy throughout the intensity enhancement. It is interesting that the strahl is broader in pitch-angle from 30° to >50° during the ~3 hrs before and after the HCS crossings. Although the strahl decreases it does not completely drop out during the ST ion intensity enhancements in between the HCS crossings, indicating that those field lines were still connected to the Sun. Figure 1d reveals a clear lack of velocity dispersion during the onsets of individual intensity enhancements seen in Figures 1a & 1b. To examine the temporal evolution of the PADs and the heavy ion energy spectra, we separate out the time intervals prior to the 1$^{st}$ crossing and after the 2$^{nd}$ crossing each into two shorter time intervals that represent regions traversed by PSP

further away from the HCS (intervals 1A & 5A) and regions closer to the HCS (intervals 1B & 5B).

3.2 Pitch-Angle Distributions

EPI-Lo has 80 apertures that sample a wide range of pitch-angles (PAs) over a hemisphere during any given time interval (McComas et al. 2016; Hill et al. 2017). However, because of changes in the direction of the magnetic field, this coverage is neither uniform and only rarely does it sample the full (180 deg) range of PAs all the time, and so we must exercise caution when interpreting the PADs. For example, the lack of intensity enhancements during interval 4 at PAs between 140°-180° is partly due to the lack of coverage at these PAs during a large fraction of the time interval which result in more than an order of magnitude (not shown) difference between the number of EPI-Lo apertures that sampled PAs between 100°-140°. Figure 2 shows the PADs, plotted as differential intensity vs. the cosine of the PAs in the four intervals described above. Only data points with at least 1 count are shown. The left panels show the PADs in the spacecraft frame, while the right panels show the corresponding distributions transformed into the bulk SW plasma frame using the anisotropic Compton-Getting transformation techniques described in the literature (see Ipavich 1974; Sanderson et al. 1985; Stevens & Hoyng 1986; Alevizos et al. 1999; Lario et al. 2005; Dalla & Balogh 2001). Note that the data in the 1/ speed spectrograms in Figure 1d and the He PADs in the right panels of Figure 2 are plotted after subtracting the non-negligible PSP velocity vector from the measured ion velocity.

Figure 2 shows the following: (1A) further away from the HCS and prior to the 1st full crossing, the PADs show anti-sunward, field-aligned flow in the PSP frame, but become nearly flat when transformed into the plasma frame. (1B) Closer to the HCS, the PADs in both frames show ~10:1 anti-sunward, field-aligned anisotropies which point to field-aligned beam-like distributions. (5A) After the 2nd crossing and closer to the HCS, the PADs show a smaller (but still ~5:1) anisotropy in the SW frame compared with the PSP frame (as expected from the Compton-Getting transformation of a steep spectrum). However, the PADs are significantly different compared to the beam-like distributions seen just before the crossing, and instead exhibit anti-sunward, field-aligned flow. (5B) Far away from the HCS after the 2nd crossing, the PADs in the PSP frame show anti-sunward, field-aligned flow that peak at ~-0.75 or at PAs of ~20°-40°, which is partly due to the uneven and non-uniform PA sampling and coverage discussed above. Like the PADs seen well before the HCS crossing in Figure 2, the PADs observed further away after the HCS crossing also become nearly flat in the SW frame, which may be a result of stronger PA scattering and a convective, rather than streaming, anisotropy. In summary, the PADs further away from the HCS appear to indicate more isotropic distributions compared with those seen closer to the HCS crossing.

### 3.3 Spectral Properties

Figure 3 shows the H, He, O, and Fe differential energy spectra in six different time intervals. Energy spectra with three or more statistically significant data points with relative uncertainty ≤100% are fitted with power-laws of the form $j = j_0 E^{-\gamma}$. Spectral slopes for energy spectra with only two statistically significant data points are calculated using $\gamma = -\ln(j_1/j_2) / \ln(E_1/E_2)$. These time-resolved spectral slopes and indices, along with the event-averaged values for He, O, and Fe are provided in Table 1. The figures and the table show that the He, O, and Fe spectra behave as steep power-laws with indices and slopes between ~4-6±0.3/1.2 in all time intervals. In the two intervals prior to the HCS crossing, the O and Fe spectra appear to be somewhat harder than the He spectra. However, just after the 2nd crossing when the He intensities reached their peaks (see Figure 1a) and also when averaged over the entire event, the spectral indices for the three species are similar within the uncertainties, with the species-averaged value being ~5.26±0.76, which is consistent with a steep velocity phase-space distribution function $f \propto v^{-12.5}$. Thus, it is plausible that the >10 keV nucleon$^{-1}$ ST ions seen near the E07 HCS crossing constitute the tail of the heated and accelerated SW distributions, as suggested by Kasper et al. (2017) and shown in Bale et al. (2021).

3.4 Maximum Observed Energy

    In order to estimate the maximum energy for each species we examine the raw triple coincidence event data from 0839-2210 UT on 2021 January 17. Figure 4a shows the energy nucleon$^{-1}$ calculated from the ion speed measured by the start and stop ToF signals plotted vs. the total incident energy for all raw events that were identified as H, He, O, or Fe. The total incident energy is the characteristic energy of the bin that a particular event is assigned to by the onboard lookup tables. The energy bin assignment is determined after estimating energy losses in the start and stop foils, the SSD dead layer, and pulse height defect, and thus represents the best available estimate for the ion's incident energy. The green diagonal lines represent species tracks that were well-defined and populated during the 2020 November 25 SEP event (see Hill et al. 2021, in preparation). The maximum observed energies for each species as determined from these triple coincidence event data are shown as vertical dotted lines and provided in Table 1 as energy nucleon$^{-1}$, total energy, energy/charge, and particle rigidity. For calculation of energy/charge and rigidity, we assume average ionization states for O6.15+ and Fe10.6+ measured in large gradual SEPs at 1 au between ~0.18-0.24 MeV nucleon$^{-1}$ (Klecker et al. 2006a, b; 2007). It is clear that these four species do not share a common maximum total energy, energy/nucleon, energy/charge, or rigidity.

4. DISCUSSION

During encounter 07 near perihelion at 0.095 au, PSP crossed the HCS at 1340 UT on 2021 January 17. FIELDS and SWEAP detected an anti-sunward directed reconnection exhaust, with the exhaust field lines disconnected from the Sun (Phan et al. 2021b). Key IS☉IS observations during this HCS crossing are as follows:

1) EPI-Lo observed significant increases in >10-100 keV nucleon$^{-1}$ He, O, and Fe intensities in the separatrices before and after the two HCS crossings, with the peak occurring just after IMF polarity reversal associated with the 2$^{nd}$ partial crossing. Interestingly, very few heavy ions are detected inside the 1$^{st}$ full HCS crossing, and the heavy ion intensities are significantly reduced during the 2$^{nd}$ partial crossing.

2) The intensity onsets or peaks exhibit no clear velocity dispersion and occur in conjunction with the broadening of the strahl PAs from 30° to >50°.

3) The He PADs when transformed into the bulk plasma frame, like the strahl, closely track the magnetic field polarity reversal associated with the HCS and show up to ~10:1 anti-sunward, field-aligned beams and/or flows closer to the HCS, but are essentially flat further away, thus potentially indicating the presence of nearly isotropic distributions.

4) The He, O, and Fe spectra behave as steep power-laws with E$^{-4-6}$ before and after the HCS crossing, with each species exhibiting a different maximum incident energy (total energy, energy/nucleon, energy/charge, and rigidity).

We now discuss implications of these results in terms of local vs. remote sources and diffusive vs. reconnection-driven particle acceleration models. For the purpose of this discussion, we distinguish between remote or near-solar sources vs. relatively local sources, defined here as those within ~1 R$_s$ of the PSP location. These new PSP observations, as summarized in Table 2 and discussed below, raise several important questions about the origin, acceleration, and propagation of ST ions around the HCS inside 0.1 au.

4.1 Velocity Dispersion, PADs, and Local vs. Remote Acceleration

The lack of clear velocity dispersion within the 1-minute time resolution of the 1/ion speed spectrograms in Figure 1c along with the absence of concomitant flares or near-sun CME shocks on January 17, 2021, as reported by Earth-based solar observatories (see https://solen.info/solar/old_reports/2021/january/20210118.html), rules out the possibility that these ST ions are accelerated and simultaneously injected over a range of energies by near-solar sources but instead points to one of three possibilities: (1) PSP entered flux tubes filled with ST ions that were accelerated remotely (e.g., Mazur et al., 2000), (2) ST ions are being accelerated by unidentified remote sources that inject ions continuously or in a quasi-steady manner, or (3) ST ions are being accelerated at local or nearby sources located within ~1 $R_s$ of PSP. The presence of anisotropic PADs with field-aligned beams and flows closer to the HCS crossings indicates weak PA scattering and is somewhat difficult to reconcile with the notion that PSP entered flux tubes that are filled with remotely accelerated ST ions, since typically such populations tend to exhibit nearly isotropic PADs (e.g., Desai et al., 2020; Cohen et al., 2020).

The temporal evolution of the He PADs observed either side of the HCS crossing also appears to be inconsistent with a remote or solar origin, since ion anisotropies in SEP events accelerated in flares or by near-Sun CME-driven shocks tend to exhibit strong anti-sunward flows during their onsets and become more isotropic later when the intensities peak, presumably due to PA scattering and/or diffusive transport (e.g., Kallerode & Wibberenz 1997; Lario et al. 2005; Giacalone et al. 2020; Mitchell et al. 2020; Mason et al. 2020). In contrast, we find nearly isotropic distributions at the start and the end of the event, which indicates relatively strong PA scattering or diffusive transport further away from the HCS, while strong (~10:1) anti-sunward beam-like distributions are observed just before the 1st crossing and anti-sunward flows just after the 2nd crossing. As discussed above, the evolution of the PADs is difficult to reconcile with the notion that PSP entered flux tubes filled with previously accelerated ST ions, and also with scenario (2) in which the ST ions are being accelerated remotely by quasi-steady sources. Instead, this behavior indicates that PSP is probably traversing regions with magnetic field lines directly connected to the acceleration site sunward of PSP either side of the HCS, but not further away. In summary, the time evolution of the PADs is remarkably different to the behavior that is seen routinely during solar or remote origin ion events, and this in combination with the absence of velocity dispersion makes it highly unlikely that these HCS-associated ST ions are accelerated by near-solar sources.

Since scenario (3) is plausible, we estimate the possible location of local sources of these ST ions as follows. The observed minimum and maximum ion energies (in energy per nucleon) are ~9.4 keV nucleon$^{-1}$ for Fe and ~91 keV for protons, which correspond to minimum and maximum speeds of ~1345 km s$^{-1}$ and ~4200 km s$^{-1}$, respectively. Thus in 60 s, the lowest speed Fe ions could travel from a source located less than ~80,000 km away, while the higher energy protons travel this distance in 19 s, possibly yielding to the observed lack of velocity dispersion. The anti-sunward flows and beams seen in the He PADs suggest that these sources are likely located sunward of PSP, while the persistence of the strahl indicates that the ST ions are injected on field lines that remain connected to the Sun. We note that the reconnection x-line is at least ~43,000 km sunward of PSP (Phan et al. 2021b; in preparation) and could thus serve as a candidate acceleration site/source for the <100 keV nucleon$^{-1}$ H, He, O, and Fe ions during the event (see Section 4.3).

4.2 Energy Spectra, Maximum Energy, PADs, and Diffusive Acceleration

Cohen et al. (2005), Mason et al. (2012b), and Desai et al. (2016 a, b) showed that energy-dependence of heavy ion abundances, time evolution of Fe/O ratios, heavy ion spectral breaks, and other key SEP properties can be explained in terms of a resonance condition where different heavy ion species, X, have the same parallel diffusion coefficient (see Li et al. 2009). Using $\kappa_{\parallel} = \frac{1}{3} v \lambda_{\parallel}$, $\lambda_{\parallel} \propto \left(\frac{Mv}{Q}\right)^{\alpha}$, and $E = \frac{1}{2} M v^2$, it follows that $\kappa_{\parallel} \sim \frac{E^{\frac{1+\alpha}{2}}}{Q^{\alpha}}$, which gives $E_X/E_H \propto (Q_X/M_X)^{\alpha}$. Under this condition, the maximum energies $E_X$ for different species when normalized to the maximum proton energy $E_H$ scale as a power-law with the ion's $Q_X/M_X$ ratio. Figure 4b shows $E_X/E_H$ vs. $Q_X/M_X$ assuming average ionization states measured at 1 au in impulsive SEPs between ~0.011-0.06 MeV nucleon[-1] (O6.67+ & Fe13.45+; DiFabio et al. 2008), in gradual SEPs between ~0.18-0.24 MeV nucleon[-1] (O6.15+ & Fe10.6+; Klecker et al. 2006a, b; 2007), and in the SW (O6.05+ & Fe9.84+; von Steiger et al. 2000). This figure shows that $E_X/E_H$ is indeed organized by a power-law of the form $(Q/M)^{\alpha}$ where $\alpha$~0.66-0.76.

The fact that each species exhibits a different maximum incident energy (total energy, energy/nucleon, energy/charge, and rigidity), particularly when considering the energy/charge also presents serious challenges for mechanisms that employ direct parallel electric fields, such as the so-called pressure cooker mechanism (e.g., Mitchell et al. 2020). It is worth noting that differences in the maximum energy for protons (~91 keV) and He (~140 keV/Q) essentially rule out a common maximum E/Q for all species during this event. This can also be seen from the significant departure of the data points from the dashed black line in Figure 4b which represents $\alpha$~1, i.e., the case where different species have the same maximum energy/charge as protons. The figure illustrates that O and Fe can potentially have the same maximum E/Q as either protons or He, but not both, and that too only if there were a highly unlikely combination of ionization states, e.g., same E/Q as protons requires Fe18.1+ & O7.1+, while the same E/Q as He requires Fe11.8+ & O4.63+. Both combinations correspond to coronal equilibrium temperatures that are more than an order of magnitude different and would essentially require that the two species originate from distinct source regions (Mazzotta et al. 1998; Mason et al. 2016), which is inconsistent with the similarities in the H, He, O, and Fe time-intensity profiles shown in Figures 1a & 1b, the power-law spectral forms shown in Figure 3, and the corresponding spectral indices listed in Table 1.

While the organization of the maximum energies with Q/M appears to hint at a resonant diffusive-type acceleration mechanism, we remark that the presence of up to ~10:1 anti-sunward field-aligned beams and flows points to weak PA scattering closer to the HCS. This may pose challenges for first-order Fermi-type diffusive acceleration and transport mechanisms which typically produce near isotropic PADs (e.g., Zank et al. 2014; le Roux et al. 2015; Du et al. 2018; Li, Guo, & Li 2019; Pezzi et al. 2021). In addition, the fact that He, O, and Fe spectra exhibit rather steep power-laws with $E^{-4.6}$ before and after the HCS crossing may also be difficult to reconcile with diffusive acceleration, which tends to produce somewhat harder spectra (e.g., Zank et al. 2014).

4.3 Energy Spectra, Maximum Energy and Scale Sizes, and Magnetic Reconnection

As alluded to in Section 4.1, a potential local acceleration site is the reconnection x-line, which was at least ~43,000 km sunward of PSP. In-situ observations over the last two decades have identified important signatures in the magnetic field and plasma data that confirm the occurrence of magnetic reconnection in a wide variety of locations such as the Earth's magnetotail (e.g., Nagai et al., 1998; Angelopoulos et al., 2004; 2008; Øieroset et al., 2001), the Earth's magnetopause (Paschmann et al., 1979; Phan et al., 2001), the solar wind near 1 au (e.g., Gosling et al. 2005a, b; Phan et al. 2006; Gosling & Phan 2013; Khabarova & Zank 2017; Khabarova et al. 2016; 2020), the solar wind beyond 1 au (Gosling et al., 2006), and more recently, the inner heliosphere traversed by PSP (Phan et al. 2020; 2021a; Lavraud et al., 2020; Froment et al., 2021).

Although Khabarova & Zank (2017) and Kabarova et al. (2020) report observations of local in-

situ reconnection-driven acceleration of energetic ions up to ~5 MeV at the L1 Lagrangian point

at the Advanced Composition Explorer (ACE), this interpretation is somewhat controversial, since

the Alfvén speed in the solar wind at 1 au is on the order of ~10s of km/s, and hence the magnetic

energy available for particle energization is considered to be insufficient for accelerating ions up

to suprathermal energies (e.g., Gosling et al. 2005c; Phan et al. 2014). In contrast, around the HCS

crossing during E07 from ~0839-2210 UT on 2021 January 17, the local Alfvén speed at PSP was

~120s km s$^{-1}$ (Phan et al. 2021b). We remark that while the higher Alfvén speeds during E07 could

provide sufficient magnetic energy to heat and accelerate solar wind protons, presently it is not

clear whether the energy budget would also be able to simultaneously account for the <100 keV

ST tails reported in this paper.

Like diffusive acceleration, reconnection-driven particle acceleration processes are also based on first- or second-order mechanisms but here the particles gain energy primarily due to the curvature drift from magnetic island contraction or merging (e.g., Zank et al. 2014; le Roux et al.

2015; Drake et al. 2009). For energization to occur, all reconnection-driven acceleration mechanisms require the particle gyroradius $\rho \ll t_{HCS}$, where $t_{HCS}$ is the thickness of the HCS. The

HCS width is estimated to be ~40,000 km or ~8500 x $d_{ion}$, where $d_{ion}$ is the ion inertial length

(Phan et al. 2021b). From the maximum energies in Table 1 and the magnetic field magnitude

measured by FIELDS before, during, and after the HCS crossing, we estimate a gyroradii range of

677 < $\rho$ < 3045 km inside the HCS and a range of 157 < $\rho$ < 555 km outside the HCS. Given that closer to the HCS, most of the ions are field aligned and thus have smaller PAs, these values provide upper limits for $\rho$. It thus appears that the maximum $\rho < t_{HCS}$, which satisfies the required scale size for reconnection-associated mechanisms to operate. In this case, the organization of the maximum energy with Q/M could be a consequence of the particle energy gain scaling with the ion mass or Q/M ratio until the corresponding gyroradius approaches the scale size of the HCS and escapes (e.g., Drake et al. 2009; Drake & Swisdak 2012; Arnold et al. 2021), rather than due to a resonant diffusive-type mechanism as found for large SEP events (e.g., Desai et al. 2016b).

Several other IS⊙IS observations also appear to be qualitatively compatible with reconnection-driven mechanisms occurring at the x-line located at least ~43,000 km sunward of PSP, namely: the lack of clear velocity dispersion during the intensity onsets and peaks, the presence of steep power-laws, and the anti-sunward field-aligned beams or flows (e.g., Drake et al. 2009; Drake, Swisdak, & Fermo 2013). Indeed, more recently, Arnold et al. (2021) proposed a solar flare associated reconnection-driven first-order Fermi-type acceleration mechanism that produces steep power-laws during solar flares with spectral indices between ~3.5-6.5 for electrons. It remains to be seen whether such mechanisms could also operate in the near-sun HCS and produce the steep ST ion power-laws reported here. Additionally, despite the qualitative agreement between expectations of reconnection-driven acceleration models and some of our observations, we remark that the total absence or significant reduction of ion intensities during the two HCS crossings, i.e., within the reconnection jets or exhausts poses a serious challenge for such models, as prior observational studies have shown that the highest intensity enhancements occur inside the diffusion region as predicted by models (e.g., Øieroset et al. 2002; Drake et al. 2009). Another important and puzzling aspect of these observations is that the ST ion intensities peak in conjunction with the persistent presence and broadening of the electron strahl either side of the HCS crossings i.e., when PSP traversed the separatrices, indicating that these field lines are still connected to the sun as opposed to the strahl dropouts seen during the HCS separatrix crossing in E08 (see Phan et al. 2021b). The strahl broadening in the separatrices may also indicate that the electrons, and indeed the ST ions, are accelerated at a common, relatively local source that is not necessarily in the immediate vicinity of PSP (see Phan et al., 2021b). Future modeling efforts that include detailed comparisons with data will need to investigate whether reconnection-driven particle energization and release could operate near the HCS in the near-Sun SW and account for the strahl and the ST ions observed just outside the exhausts in the separatrix regions traversed by PSP, as reported in this paper.

5. SUMMARY & CONCLUSIONS

In summary, the lack of velocity dispersion and the temporal evolution of the He PADs rules out remote acceleration sites and processes such as flares and near-Sun CME-driven shocks. The fact that the maximum energies for the four ion species do not occur at the same E/q poses serious challenges for the pressure cooker mechanism. Instead, the maximum energies are $\propto$ $(Q_X/M_X)^\alpha$ where $\alpha \sim 0.7$ which, in turn, points to either a resonant diffusive-type mechanism or mechanisms in which particles gain sufficient energy, increase their gyroradii, and escape from the acceleration region. Among the two candidates for local acceleration, our observations present the most serious challenges for diffusive mechanisms as these tend to produce harder energy spectra and nearly isotropic PADs. Possible sources for the ST ions near the 2021 January 17 HCS crossing may involve contracting magnetic islands near the reconnection x-line located $\sim 2 \times 10^5$ km sunward of PSP. However, currently even such magnetic reconnection-based acceleration models are unable to account for the total or near absence of ST ions inside the two HCS crossings which comprise the reconnection jets or exhausts. This highlights the need to re-evaluate potential sources of these near-solar, current sheet associated superathermal particles. In conclusion, no single existing model or theory appears to fully account for all of the ISʘIS/EPI-Lo observations presented in this paper, thus requiring a re-examination of existing theories of ST ion production close to the Sun.

Acknowledgements

This work was supported by NASA's Parker Solar Probe Mission, contract NNN06AA01C. We thank all the scientists and engineers who have worked hard to make PSP a successful mission, in particular the engineers, scientists, and administrators who designed and built ISʘIS/EPI-Lo, ISʘIS/EPI-Hi, FIELDS, and SWEAP instrument suites, and support their operations and the scientific analysis of its data. For their contributions to the scientific configuration and instrumental analysis, we owe special thanks to P. Kollmann, J. Peachy, and J. Vandegriff at JHU/APL for EPI-Lo. The ISʘIS data are available at http://spp-isois.sr.unh.edu/data_public/ as well as at the NASA Space Physics Data Facility. Work at SwRI is supported in part under NASA grants 80NSSC20K1815, 80NSSC18K1446, 80NSSC21K0112, 80NSSC20K1255, and 80NSSC21K0971.

Table 1: Time-resolved and event-averaged spectral indices of He, O, and Fe measured by IS☉IS /EPI-Lo during the E07 HCS crossing. The different time intervals are identified in Figure 1d. Maximum incident energy for each species is determined from triple coincidence events analyzed by IS☉IS/EPI-Lo from 0839-2210 UT on 2021 January 17 (see Figure 4a).

| Species | Energy Range (keV n$^{-1}$) | Spectral Index, γ | | | | | Maximum Observed Energies | | | |
| | | 1A: Prior to HCS Crossing 1: 0829-1119 | 1B: Prior to HCS Crossing 1: 1122-1315 | 5A; After HCS Crossing 2: 1443-1710 | 5B: After HCS Crossing 2: 1442-1900 | Event Averaged: 0839-2210 | E/nucleon (keV n$^{-1}$) | E (keV) | E/Q (keV/ Q)$^a$ | Rigidity (MV)$^a$ |
|---|---|---|---|---|---|---|---|---|---|---|
| H | ... | ... | ... | ... | ... | ... | 91 | 91 | 91 | 13 |
| $^4$He | 30-90 | 5.94 ± 0.70 | 5.83 ± 0.74 | 5.72 ± 0.31 | 6.62 ± 0.74 | 6.07 ± 0.33 | 70 | 281 | 140 | 23 |
| O | 15-75 | 3.63 ± 0.31 | 3.40 ± 0.37 | 4.24 ± 0.49 | 5.90 ± 0.55 | 4.56 ± 0.21 | 40 | 647 | 105 | 23 |
| Fe | 20-40 | 4.89 ± 1.18 | 4.53 ± 1.12 | 4.69 ± 0.63 | 6.42 ± 1.20 | 5.14 ± 0.82 | 29 | 1648 | 156 | 39 |

Notes

[a]E/Q and ion rigidity, R, are estimated using average ionization states measured between ~0.18-0.24 MeV nucleon$^{-1}$ in large gradual solar energetic particle events (LSEPs) at 1 au (Klecker et al. 2007).

Table 2: Qualitative comparison between key ISOIS observations during the January 17, 2021 heliospheric current sheet crossing event and model/theoretical predictions.

| PSP/ISOIS/EPI-Lo Observations | Properties/Implications | Local | | Remote | |
|---|---|---|---|---|---|
| | | Diffusive models | Reconnection models | Flare or CME-associated | Direct $E_{\parallel}$ Acceleration |
| Intensity peaks occur near the HCS crossings but outside the reconnection exhausts | No association with local shocks, compression regions, or visible solar activity | ✓ | X | X | ✓ |
| No velocity Dispersion | Combined with time evolution of PADs, most likely points to local or nearby sources | ✓ | ? | X | X |
| Isotropic PADs far away from HCS crossings | Strong PA scattering, convective anisotropy | ✓ | ? | X | ? |
| Up to ~10:1 anti-sunward, field-aligned beams and flows closer to the HCS crossings | Weak PA scattering with direct connection to the source region sunward of PSP | X | ✓ | ✓ | ✓ |
| Spectra are steep power-laws with $j \propto E^{-5.26 \pm 0.76}$ | Maxwellian-like phase space density distribution with $f \propto v^{-12.5}$. Possible ST tail of SW distribution. | X | ✓ | ✓ | ✓ |
| Maximum Observed Energy $E_X$ for species X | Not ordered by E, E/nucleon, E/Q or R. $E_X/E_H \propto (Q_X/M_X)^{0.65-0.76}$ | ✓ | ✓ | ✓ | X |

Figures and Figure Captions

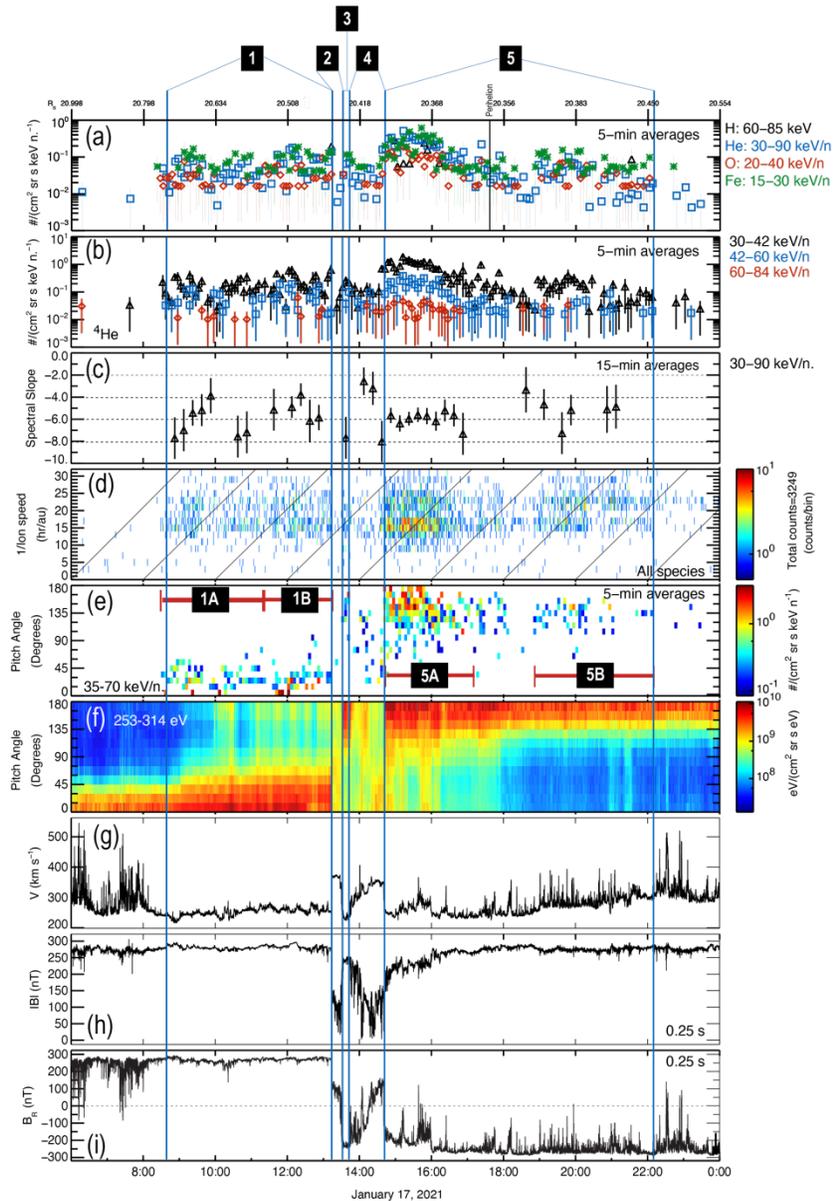

Figure 1: (a) 5-minute averages of ~15-90 keV nucleon⁻¹ H, He, O & Fe intensities, (b) 5-minute averages of ~30-85 keV nucleon⁻¹ He intensities, (c) 15-minute averages of ~30-90 keV nucleon⁻¹ He spectral slope, (d) 1-minute averages of the 1/ion speed spectrograms constructed from all valid triple coincidence ion counts (ToF vs. E) detected by EPI-Lo, (e) 5-minute averages of ~35-70 keV nucleon⁻¹ He pitch-angle distributions (PADs) in the spacecraft frame, (f) 0.87 s averages of ~253-314 eV electron PADs, (g) ~3.5 s average of the bulk solar wind speed, and (h) ~0.25 s averages of magnetic field magnitude, and (i) the radial component of the magnetic field in the sun-centered RTN coordinate system. Black vertical bar in (a) shows PSP perihelion. Vertical blue bars: time intervals identified as: (1) prior to 1st crossing, (2) during 1st crossing, (3) in between two crossings, (4) 2nd partial crossing, and (5) after 2nd crossing. Time periods 1A, 1B, 5A & 5B in (e) identify time intervals selected for analyses of PADs and energy spectra (see text for details).

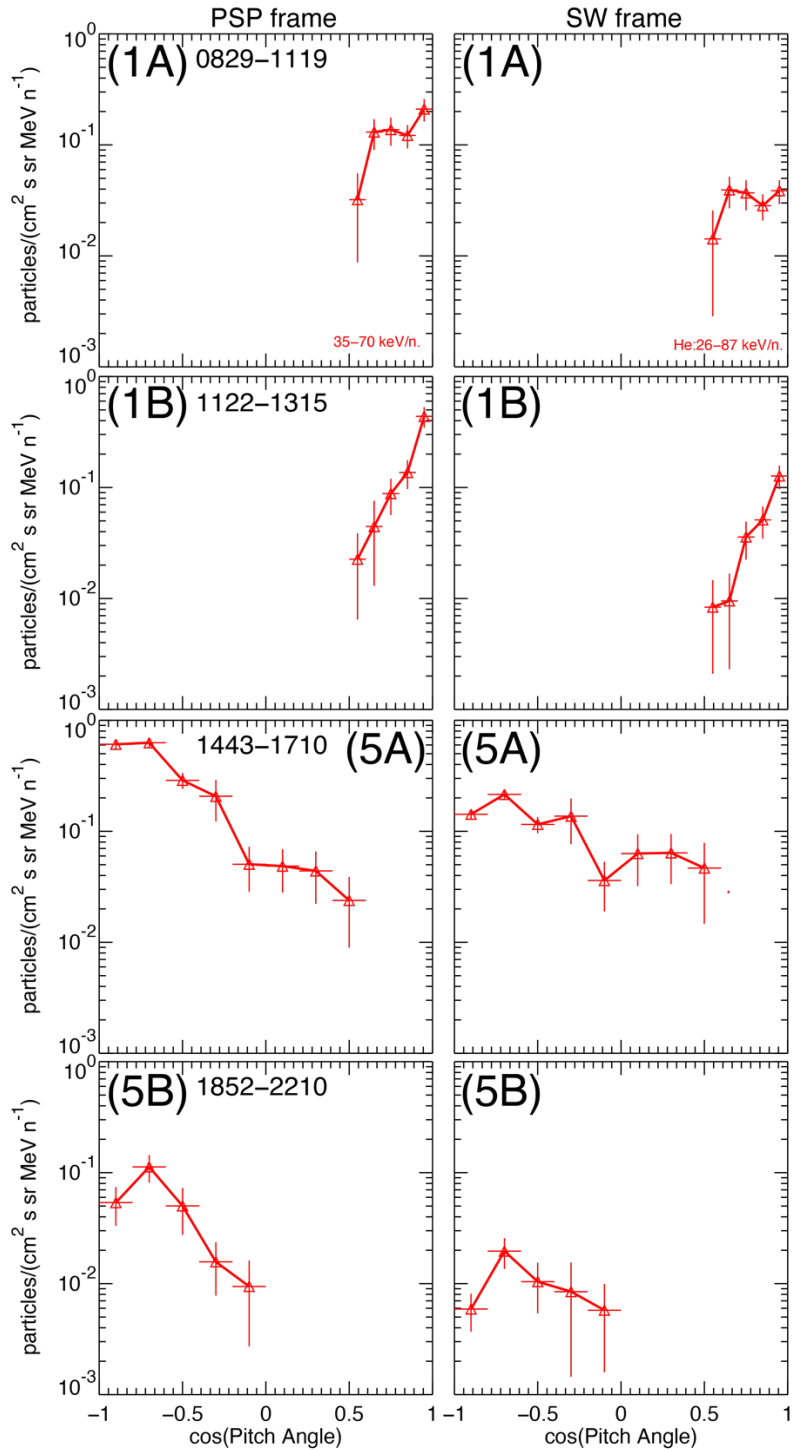

Figure 2: PADs plotted as differential intensity vs. cosine of the PAs averaged during the four time intervals identified in Figure 1e. Left: PADs in spacecraft frame. Right: PADs transformed into the bulk solar wind plasma frame.

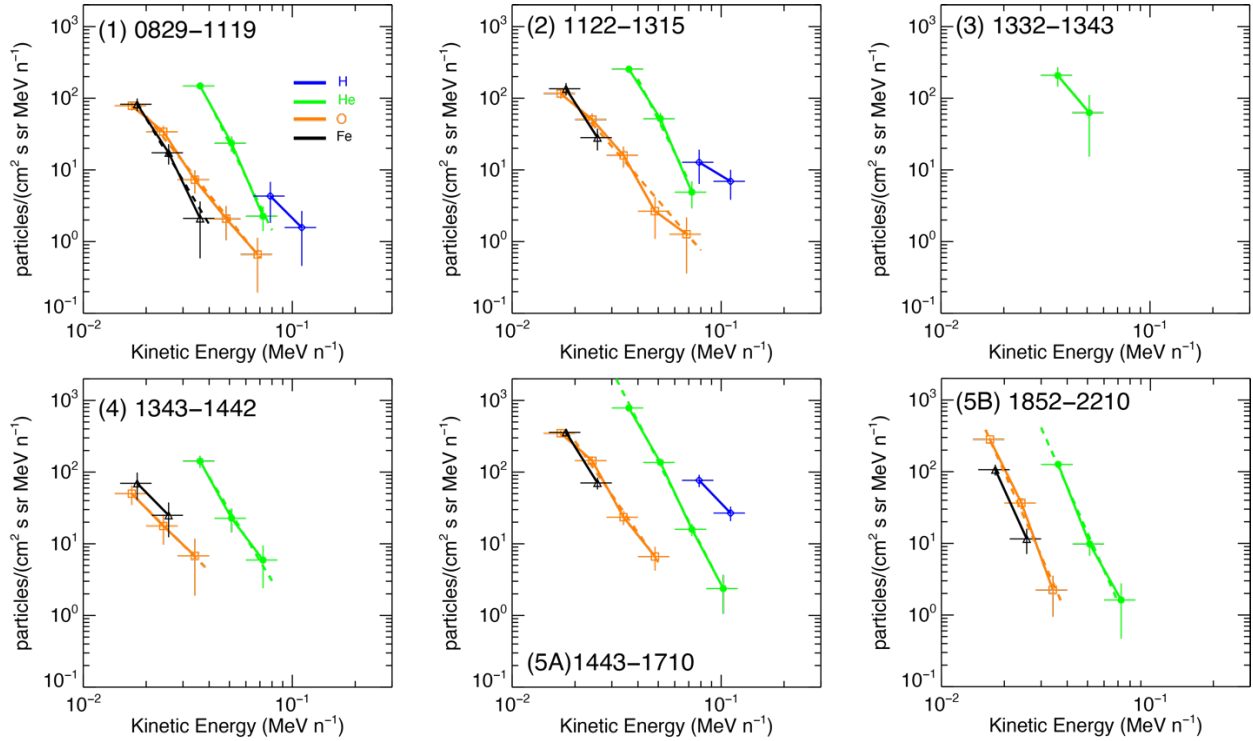

Figure 3: Differential energy spectra for H, He, O, and Fe in six different time intervals associated with the HCS crossing (see Figure 1). Dashed lines represent power-law fits to the He (green), O (orange), and Fe (black) spectra, with the power-law spectral indices provided in Table 1.

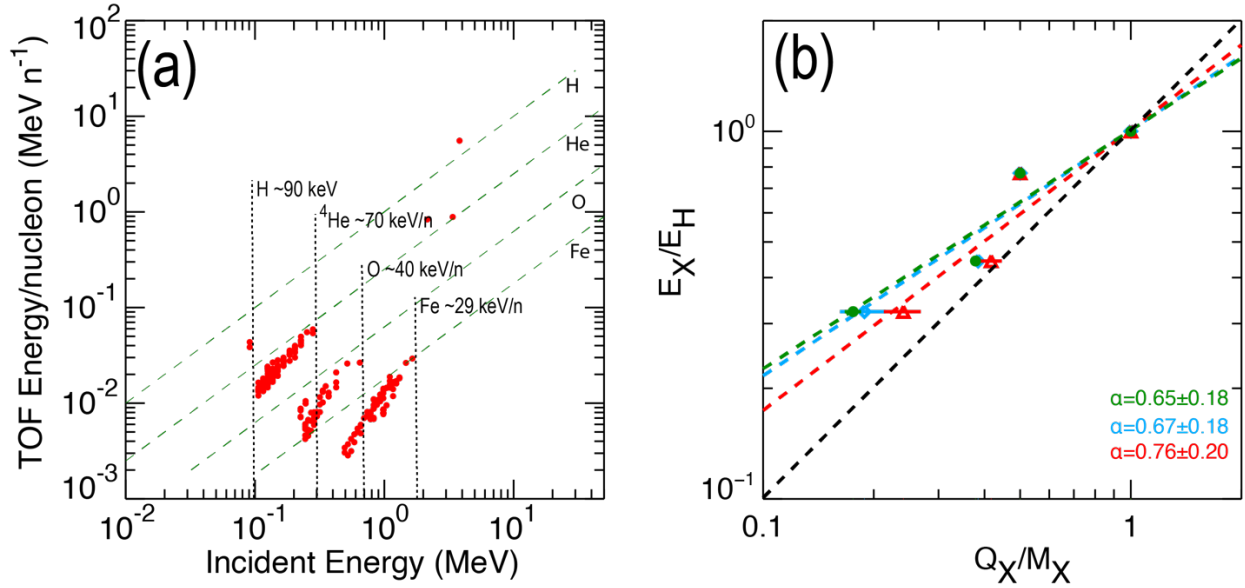

Figure 4: (a) Energy/nucleon (MeV/nucleon) estimated from the Time of flight (ToF) measurement vs. the total incident energy (MeV) of all triple coincidence H, He, O and Fe events identified by IS⊙IS/EPI-Lo from 0839-2210 UT on 2021 January 17. The ToF energy/nucleon is determined from the ion speed or ToF measurement and the total incident energy is estimated from the residual energy deposited in the SSD after accounting for energy losses in the start and stop foils, the SSD dead layer, and pulse height defect (see Hill et al. 2021 in preparation). Green diagonal lines show the location of H, He, O and Fe tracks as determined from similar raw event data obtained during the high count rate 2020 November 20 CME/ESP event. Dotted vertical lines show the maximum incident energy for each species, as determined from the corresponding total incident energy. (b) Maximum observed energy $E_X$ versus ion charge-to-mass ratio for that species. O and Fe ionization states are assumed to be event-averaged charges states measured in solar wind (green), impulsive (red), and gradual (blue) SEPs. Colored dashed lines represent the corresponding fits to $E_X/E_H \propto (Q_X/M_X)^{\alpha}$; dashed black line represents $E_X/E_H \propto (Q_X/M_X)$, i.e., $\alpha = 1$.